\documentclass[12pt,english]{article}
\usepackage[letterpaper]{geometry}
\usepackage[font=small, width=0.9\textwidth]{caption}
\usepackage{amsmath}
\usepackage{amsthm}
\usepackage{amssymb}
\usepackage{authblk}
\usepackage{babel}
\usepackage{graphicx}
\usepackage{hyperref}
\usepackage{lmodern}
\usepackage{mathrsfs}
\usepackage{sistyle}
\usepackage{subfigure}  
\usepackage{verbatim}   
\usepackage{upgreek}
\usepackage{xcolor}
\usepackage{comment}
\usepackage{cprotect}
\makeatletter
\usepackage[numbers,sort&compress,comma]{natbib}
\usepackage{hyperref}
\hypersetup{
	colorlinks=true,
	linkcolor=blue,
	urlcolor=red,
	citecolor=blue
}
\usepackage{fancyhdr}
\usepackage{enumitem}
\setenumerate{itemsep=5pt,topsep=5pt}

\usepackage{footmisc}

\renewcommand\section{%
	\@startsection {section}{1}%
	{\z@}{-16pt \@plus 0pt \@minus 0pt}%
	{8pt \@plus.2ex}%
	{\normalfont\large\bfseries\scshape}}

\renewcommand\subsection{%
	\@startsection {subsection}{1}%
	{\z@}{-8pt \@plus 0pt \@minus 0pt}%
	{4pt \@plus.2ex}%
	{\normalfont\normalsize\bfseries}}

\renewcommand\subsubsection{%
	\@startsection {subsubsection}{1}%
	{\z@}{-8pt}%
	{0.001pt \@plus 0pt}%
	{\normalfont\normalsize\bfseries}}

\renewcommand{\paragraph}{%
	\@startsection{paragraph}{4}%
	{\z@}{-8pt}{-0.5em}%
	{\normalfont\normalsize\bfseries}%
}

\makeatother

\newcommand{\um}{\upmu\text{m}}

\newcommand{\SiO}{SiO$_2$}
\newcommand{\TaO}{Ta$_2$O$_5$}

\newcommand{\tantalum}{$\text{Ta}_\text{2}\text{O}_{\text{5}}$~}
\newcommand{\oxide}{$\text{SiO}_{\text{2}}$~}
\newcommand{\nitric}{$\text{NHO}_{\text{3}}$~}

\begin{document}
	
	\title{Monitored wet-etch removal of individual dielectric layers from high-finesse Bragg mirrors}
	
	\author{Simon Bernard\footnote{These authors contributed equally}, Thomas J.~Clark$^*$, Vincent Dumont, Jiaxing Ma, Jack C. Sankey\footnote{jack.sankey@mcgill.ca}}
	
	
	\affil{McGill University Department of Physics\\
		3600 rue University, Montr\'{e}al QC, H3A 2T8, Canada}
	
	\maketitle
	
	
	\begin{abstract}
		It is prohibitively expensive to deposit customized dielectric coatings on individual optics. One solution is to batch-coat many optics with extra dielectric layers, then remove layers from individual optics as needed. Here we present a low-cost, single-step, monitored wet etch technique for reliably removing (or partially removing) individual \SiO~and \TaO~dielectric layers, in this case from a high-reflectivity fiber mirror. By immersing in acid and monitoring off-band reflected light, we show it is straightforward to iteratively (or continuously) remove six bilayers. At each stage, we characterize the coating performance with a Fabry-P\'{e}rot cavity, observing the expected stepwise decrease in finesse from $92,000 \pm 3,000$ to $3,950 \pm 50$, finding no evidence of added optical losses. The etch also removes the fiber's sidewall coating after a single bilayer, and, after six bilayers, confines the remaining coating to a $\sim$50-$\um$-diameter pedestal at the center of the fiber tip. Vapor etching \emph{above} the solution produces a tapered ``pool cue'' cladding profile, reducing the fiber diameter (nominally 125~$\um$) to $\sim$100~$\um$ at an angle of $\sim$0.3$^\circ$ near the tip. Finally, we note that the data generated by this technique provides a sensitive estimate of the layers' optical depths. This technique could be readily adapted to free-space optics and other coatings.
	\end{abstract}
	

	\section{Introduction}\label{sec:intro} 
	
	Dielectric optical coatings represent a ubiquitous enabling technology throughout the field of optics \cite{Macleod2017Thin}, helping manage surface reflections, filter light, adjust polarization, and split or combine beams. In particular, the lowest-loss mirror surfaces are realized with dielectric Bragg stacks, allowing for the creation of etalons and other Fabry-P\'erot resonators useful as narrowband filters, wavemeters, spectrometers, and laser cavities used throughout science and engineering. These have proven invaluable for enhancing light-matter interactions, performing spectroscopy, and filtering light within the fields of cavity quantum electrodynamics \cite{Walther2006Cavity}, quantum information \cite{Reiserer2015Cavity} (including solid-state systems, e.g. \cite{Riedel2017Deterministic}), optomechanics \cite{Aspelmeyer2014Cavity}, optical-wavelength astronomy \cite{Haffner2009The}, and gravitational wave detection \cite{Abbott2016Observation}.
	
	Within these (and other) contexts, one routinely wishes to choose a specific mirror reflectivity, for example to make a trade-off between the storage time of a cavity and collection efficiency. Currently, however, it is expensive to deposit a customized, high-quality coating on a small number of optics, as this requires a dedicated deposition chamber. An alternative approach is to deposit more dielectric layers than are required upon \emph{many} substrates, and then remove them as needed to tune the coating properties. There exists a great body of knowledge about different etch processes and materials \cite{Williams2003Etch}, but when it comes to high-finesse mirror coatings, it is not known how these will introduce additional optical losses through surface roughness and contamination. Low-loss removal of dielectric mirror coatings has been achieved using plasma etching \cite{Purdy2008Integrating, Purdy2009Cavity}, however, this technique relies on a cumbersome reactive ion etching machine, careful control over electric field and gas distribution within the chamber, and periodically recalibrated etch rates. Also, typical chambers introduce complications when attempting to etch a substrate (e.g., a fiber mirror \cite{Hunger2010A}) having a geometry significantly different from that of a silicon wafer .
	
	Here, we present a simple, monitored wet-etch technique for removing (or partially removing) dielectric layers from a high-reflectivity, low-loss \SiO/\tantalum Bragg mirror. We show that it is straightforward to iteratively remove 6 bilayers, and, importantly, find no evidence of additional optical losses associated with the process in a finesse-50,000 Fabry-P\'erot cavity. The studied mirror coatings reside at the tips of cleaved, laser-machined optical fibers \cite{Hunger2010A}, but the etch technique should work just as well with free space optics and other \SiO/\tantalum coatings.

	\section{Monitored Dielectric Etch \& Characterization} \label{sec:measurement} 
	
\begin{figure}[!ht] 
		\centering
		\includegraphics[width=13cm]{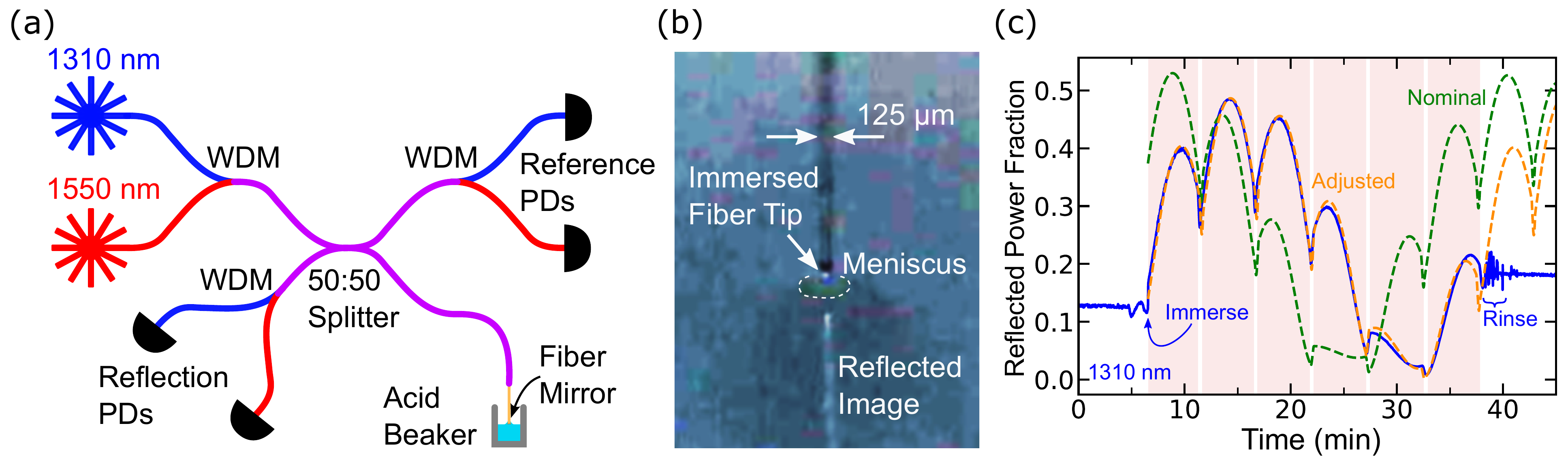}
		\caption{Etch technique. (a) Optical circuit. ``Off-band'' light ($1310$~nm, blue) and stopband light ($1550$~nm, red) is combined with a wavelength division multiplexer (WDM) and split (50:50) between ``reference'' photodiodes (PD) and the fiber mirror. The mirror is immersed in acid (Transene Co. Tantalum Etchant 111 \nitric 25-40\% by weight, HF 5-15\%, deionized water 35-70\%), and reflected 1310-nm light provides an interferometric record of the etch. (b) Fiber mirror touching the solution. A meniscus indicates the moment of contact. (c) Measured reflected off-band power, normalized by the 10 mW incident on the fiber mirror (blue) during a continuous etch. Small oscillations prior to immersion at 6.5 min arise from interference with the liquid surface. Once immersed, etching of the \tantalum layer begins at a rate of $\sim 40$ nm/min, evidenced by a slowly evolving segment of a sinusoid until $12$ min. The \oxide layer then etches $\sim 20$ times faster, appearing as a rapidly evolving sinusoid, and the process repeats, shaded regions highlight \tantalum etching. Fluctuations after 38 min correspond to interference from rinsing the fiber. A total of six bilayers are etched. The green dashed curve shows the predicted behavior using the coating company specifications and refractive indices 2.0893 for \TaO~and 1.4662 for \SiO~at 1310 nm \cite{Gao2012Exploitation} (via refractiveindex.info), and the orange dashed curve shows the prediction when these are increased by $0.7\%$ and $1.2\%$, respectively. Note the raw data are vertically scaled by 1.12 to match the model, consistent with the $\sim 10$\% uncertainty in our circuit's overall optical losses.}
		\label{fig1}
	\end{figure}

	Figure \ref{fig1}(a) shows the apparatus used to reliably remove dielectric layers from our optical coatings (Laseroptik GmbH). In this case, a high-reflectivity, low-loss Bragg mirror comprising fifteen \TaO/\SiO~bilayers (terminated with \TaO) on the end of a cleaved, laser-ablated \cite{Hunger2010A} Corning SMF-28 fiber is immersed in Tantalum Etchant 111 (Transene Co.; \nitric 25-40\% by weight, HF 5-15\%, deionized water 35-70\%), which, as we estimate here, provides a convenient 20:1 etch rate ratio for \SiO:\TaO. Our stack's stopband nominally spans 1420-1710~nm and the power reflectivity $|r|^2\approx 0.99997$ at 1550~nm. Figure \ref{fig1}(b) shows an image from a low-cost USB microscope, highlighting the visible meniscus that appears the moment the tip is immersed and etching begins.
	
	During the etch process, we monitor ``off-band'' (1310 nm) light reflected from the coating, which provides a high-contrast interferometric signal (blue data in Fig.~\ref{fig1}(c)) exhibiting distinctive features associated with each layer. During \TaO~etches ($40$~nm/min), we observe a slowly evolving fringe (e.g., minutes 7-12), and during the \SiO~etch ($800$~nm/min), we observe a rapidly evolving fringe (e.g., the cusp near minute 12). Between layers, the sudden change in the etch rate appears as a discontinuity in the fringe slope. We also observe comparatively small fringes from interference between reflections from the fiber tip and liquid surface before and after immersion. To stop the etch, we withdraw the fiber tip while running deionized water down its length; and then (without allowing it to dry) immerse it in DI water, and finish with an isopropyl rinse (this produces additional ``noise'' in our monitor).
	
	In Figure ~\ref{fig1}(c) we also compare our measurements to a one-dimensional transfer matrix model \cite{Reinhardt2017Ultralow-Noise}. Using layer parameters specified by the coating company (dashed green), the agreement is poor, but if we increase the optical depths of \tantalum and \oxide by $0.7\%$ and $1.2\%$ respectively, the model prediction changes dramatically, and the agreement is compelling. In addition to building confidence in our interpretation, this exercise provides a refined model of our coating for other experiments involving these mirrors. 
	
	As shown in Fig.~\ref{fig1}(a), we simultaneously monitor reflected stopband light, which provides an indication that the mirror is still intact. Mostly this is a relic from initial attempts with HF solutions that destroyed the fibers before etching the first \TaO~layer; if the coating falls off, this signal suddenly drops to few percent of its original value, an event that is not as obvious in the monitored off-band light.
	
	\begin{figure}[!ht] 
		\centering
		\includegraphics[width=13cm]{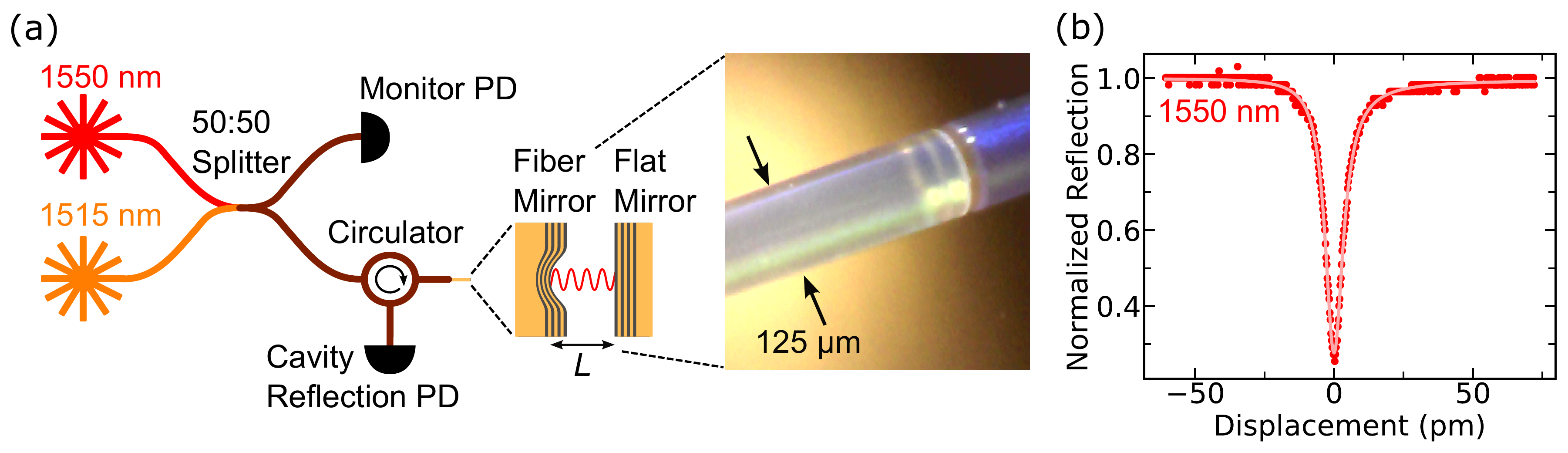}
		\caption{Fabry-P\'{e}rot characterization. (a) ``Vernier'' optical circuit used to reproducibly position a fiber mirror near a flat mirror and measure cavity finesse. Light from 1550 nm and 1515 nm sources is combined at a 50:50 splitter and sent through a circulator to the mirror. Fiber paddles (not shown) between the splitter and circulator select a single polarization mode of the cavity. Reflected light collected by a photodiode (PD) exhibits a dip (see (b)) when the cavity length is swept through resonance with either wavelength. The cavity is resonant with \emph{both} wavelengths when $L=33.3$~$\um$, and from that location we count $1550$~nm resonances until $L=6.2~\um$, where the finesse saturates. The inset shows an unetched fiber (and its reflection). (b) Reflected power at 1550~nm (symbols) normalized by the off-resonant value (0.86 mW incident on the fiber mirror, 0.43 mW collected) and fit to a Fano function (solid curve) \cite{Janitz2015Fabry} used to extract a resonance linewidth of $8.3 \pm 0.5$~pm, corresponding to finesse of $93,400 \pm 500$; statistical errors here are limited by sweep-to-sweep variations due to stage vibrations. Repeatedly mounting the fiber mirror causes the measured finesse to vary by $\pm 3000$, likely due to the cavity mode probing different regions of the flat mirror \cite{Benedikter2019Transverse}.}
		\label{fig2}
	\end{figure} 
	
	We characterize the mirror performance by measuring the finesse of a Fabry-P\'{e}rot cavity comprising the etched fiber mirror and a flat mirror having the same (unetched) coating. Our fiber dimples have radius of curvature $\sim 300~\um$, and effective mirror diameters $\sim 30~\um$, causing detrimental ``clipping'' losses when the cavity mode diameter is too large at the fiber mirror surface \cite{Benedikter2015Transverse}. For the cavity finesse to be limited by coating properties, the cavity length $L$ should be smaller than $10~\um$ in our case. To achieve this, we employ a dual-wavelength ``Vernier'' system (Fig.~\ref{fig2}(a)) capable of repeatably creating microns-long cavities of known $L$. We shine two lasers having different wavelengths (1550~nm and 1515~nm) within the coating stopband, and monitor the reflected power while varying cavity length. When the cavity is resonant with either wavelength, we observe a dip (``typical'' sweep in Fig.~\ref{fig2}(b)). Both lasers are resonant when $L=33.3~\um$ (overlapping dips), and from there we reduce $L$ while counting 1550~nm resonances to achieve the desired length. Throughout this process, $L$ is continuously swept using the flat mirror's mount (POLARIS-K1S3P), which has integrated piezoelectric actuators; we recommend a sinusoidal waveform ($\sim 20$~Hz for us), which minimizes impulse-driven vibrations. 
	
	The fiber mirror resides in a clamp mounted upon a rigid single-axis stage (Luminos model I1000). Once the desired cavity length ($L=6.2~\um$) is achieved, we tune the flat mirror angle by hand to maximize the finesse. After each adjustment, which also changes the cavity length, $L$ is returned to $6.2~\um$ using the fiber mirror's stage. It is critical to iteratively repeat this procedure until the maximum finesse is achieved, or else the results are neither reproducible nor well-matched with theory \cite{PythonCode}.
	
	Ultimately, we find a nominal cavity finesse of $92,000 \pm 3,000$ at $L=6.2~\um$, consistent with the coating company's specified 25~ppm transmission and 10~ppm losses. 
	
	\section{Iteratively Tuned Coating} \label{sec:results}
	
\begin{figure}[!ht]
		\centering
		\includegraphics[width=10cm]{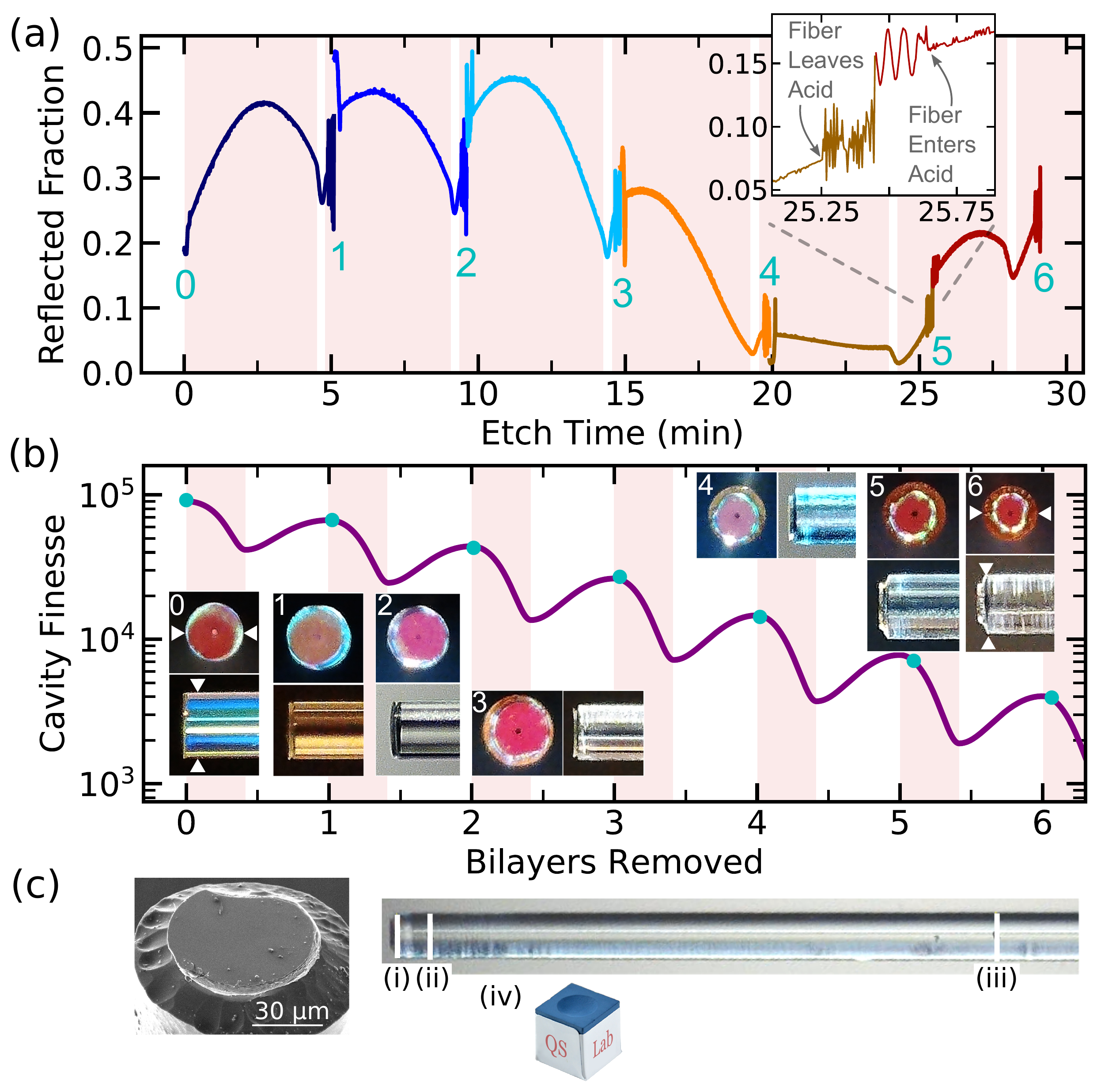}
		\caption{Iteratively tuning finesse. (a) Normalized reflected off-band power for six bilayer etches. Each etch is stopped shortly after the \SiO~layer is removed. Integers (0-6) indicate the number of bilayers removed. Inset shows the ``typical'' interference associated with removing, rinsing, and submerging the fiber. (b) Measured (teal) and predicted (violet) finesse as a function of bilayers removed. Uncertainties are smaller than the symbols, but all measurements lie within two standard deviations of the model, which includes no losses introduced by the etch. Inset images show the fiber tip at each stage. Initially, the diameter (white arrowheads in image 0) is 125~$\um$~plus a few microns of ``candy'' coating on the sidewall that is completely removed with the first bilayer etch. Etching further narrows the fiber and confines the mirror surface to a pedestal of reduced diameter. (c) Electron microscope image of fiber tip (left) after the sixth bilayer is removed. Right-hand image shows the tapered etch profile resulting from vapor etching above the solution, with diameters (i) 95~$\um$, (ii) 98.5~$\um$ after 75~$\um$, and (iii) $105$~$\um$ a distance 1.35 mm from the tip. At the expense of achievable cavity finesse, chalk (iv) may be optionally applied to increase tip traction during impact. Shaded regions in (a) and (b) highlight \tantalum etching.
			\label{fig3}}
	\end{figure}

	We now show it is straightforward to iteratively remove multiple bilayers without introducing significant optical losses. Figure~\ref{fig3}(a) shows the reflected off-band power measured over the course of six separate bilayer etches, with the time axis roughly corresponding to the total time in or near the etchant. Within seconds of finishing a \SiO~layer, we stop the etch in order to measure finesse.
	
	The teal symbols in Fig.~\ref{fig3}(b) show the finesse measured after each etch. They do not lie exactly on integer numbers of bilayers removed (x-axis) because we erred on the side of slightly over-etching somewhat into the \tantalum layer. Under certain conditions (e.g., between bilayers 5 and 6) it is more difficult to identify the transition to a new material; if this is an important issue, it can be mitigated with a second off-band monitor wavelength. The uncertainties associated with our estimates of layer transition points, etch rates, and finesse are smaller than the symbols in Fig.~\ref{fig3}(b). The violet curve shows the finesse predicted using a 1D transfer matrix model, assuming 10 ppm losses (modeled as a loss layer at the mirror surface \cite{Stambaugh2015From}) and layer thicknesses specified by the coating company; in this case, modifying the optical depths by a few percent does not significantly impact this result. 
	
	Inset optical images in Fig.~\ref{fig3}(b) show the fiber tip at each stage. Initially, the fiber diameter is $125~\um$, plus a few microns of ``candy'' coating originally deposited on the sidewall. The first bilayer etch fully removes it, and the useful mirror surface is confined to a $\sim$100-$\um$-wide pedestal. As more bilayers are removed, the pedestal becomes narrower and taller (an eventual limit on the number of bilayers that can be removed from a fiber mirror), while the cladding diameter shrinks due to sidewall etching from the solution and vapor above it. After six bilayers are removed, the fiber exhibits a tapered profile, widening from 95~$\um$ at the tip to its nominal value over a scale of millimeters. The bottom electron microscope image shows the final fiber end in more detail. The surface appears approximately flat with no obvious edge bur, allowing for the creation of microns-long cavities.

	\section{Conclusions} \label{sec:conclusion}
	We presented a simple, cost-effective etching technique for removing individual dielectric layers from a \TaO/\SiO~ dielectric coating. Importantly, this technique introduces no observable optical losses to a fiber mirror, as measured in a Fabry-P\'{e}rot cavity up to finesse $\sim$50,000. For the specific case of fiber mirrors, vapor etching above the solution produces a ``pool cue'' profile that we suspect can be tuned with a combination of surfactants (e.g., 3M Novec 4200 Electronic Surfactant), controlled gas flow, and / or protective layers applied to the cladding. Additionally, we find that a single bilayer etch eliminates the unwanted sidewall coating, which may facilitate their use in tight-tolerance ferrules. 
	
	We find that etchants involving \emph{only} HF destroy the cladding before the first \tantalum layer is removed, but, due to this high selectivity, these solutions could be used as a finishing step to precisely end the etch at the \TaO~surface, as in Ref.~\cite{Purdy2009Cavity}. We furthermore expect one can achieve selectivities between this extreme and the 1:20 by diluting the Ta etchant with an HF solution.
	
	The above measurement system is sufficient for our needs, but we can suggest a few improvements. First, multiple off-band wavelengths (in the ultimate limit, white light illumination with a spectrometer) would improve the ability to detect layer transitions. Second, we note that lasers are not strictly required, especially for coatings having only a few layers; a reasonably coherent LED source may suffice.
	
	Finally, we reiterate the sensitivity of the off-band fringes to small changes in the optical depths of the layers, as shown in Fig.~\ref{fig1}(c), and suggest that performing this simple etch on a sacrificial optic may also serve to provide valuable calibration data for a given coating run.

	\section*{Funding}
	NSERC (RGPIN 2018-05635); CFI (228130,36423); CRC (235060); INTRIQ; Centre for the Physics of Materials at McGill. 
	
	\section*{Acknowledgments}
	We thank Zhao Lu and Alireza H. Mesgar for useful discussions regarding etchants and safety protocols. SB and VD acknowledge support from FRQNT-B2 scholarships.
	
	\section*{Disclosures}
	\noindent The authors declare no conflicts of interest.

	
	\addcontentsline{toc}{section}{References}
	\bibliographystyle{bibstyle-jack}
	\bibliography{errthing}

\begin{thebibliography}{18}
\bibitem [{Pyt()}]{PythonCode}%
  \BibitemOpen
  \href@noop {} {}\bibinfo {note} {We suggest incorporating a real-time peak
  fitter into the acquisition system (Python code available upon
  request).}\BibitemShut {Stop}%
\end{thebibliography}
	
\end{document}